\documentclass[number,sort&compress,twocolumn,3p]{elsarticle}
\usepackage[binary-units=true,separate-uncertainty=true]{siunitx}
\usepackage{lineno,hyperref}
\usepackage{tikz}
\usepackage{amssymb,amsmath,mathtools}
\usepackage{caption}
\usepackage{xspace}
\usepackage{booktabs}
\usepackage{siunitx}
\usepackage{textcomp}
\usepackage[percent]{overpic}
\usepackage[list=true,listformat=simple,margin=10pt]{subcaption}
\usepackage{hepnames}

\journal{Nucl. Instr. Meth. A}

\newcommand{\apsq}{\texorpdfstring{\ensuremath{\mathrm{Allpix}^2}}{Allpix\textasciicircum 2}\xspace}
\newcommand{\corry}{Corryvreckan\xspace}

\begin{document}
\begin{frontmatter}
  \title{Allpix$^2$: A Modular Simulation Framework for Silicon Detectors}

\author[cern]{S.~Spannagel\corref{corr}}
\ead{simon.spannagel@cern.ch}
\author[cern]{K.~Wolters\fnref{eth}}
\author[cern]{D.~Hynds\fnref{nikhef}}
\author[cern]{N.~Alipour Tehrani}
\author[unige]{M.~Benoit}
\author[cern]{D.~Dannheim}
\author[unige]{N.~Gauvin}
\author[cern]{A.~N\"urnberg\fnref{kit}}
\author[desy]{P.~Sch\"utze}
\author[unige]{M.~Vicente}

\address[cern]{CERN, Geneva, Switzerland}
\address[unige]{Universit\'e de Gen\`eve, Geneva, Switzerland}
\address[desy]{DESY, Hamburg, Germany}

\cortext[corr]{Corresponding author}
\fntext[eth]{Now at ETH Zurich, Switzerland}
\fntext[nikhef]{Now at Nikhef, Amsterdam, Netherlands}
\fntext[kit]{Now at KIT, Karlsruhe, Germany}

\begin{abstract}
  \apsq (read: Allpix Squared) is a generic, open-source software framework for the simulation of silicon pixel detectors.
  Its goal is to ease the implementation of detailed simulations for both single detectors and more complex setups such as beam telescopes from incident radiation to the digitised detector response.
  Predefined detector types can be automatically constructed from simple model files describing the detector parameters.

  The simulation chain is arranged with the help of intuitive configuration files and an extensible system of modules, which implement separate simulation steps such as realistic charge carrier deposition with the Geant4 toolkit or propagation of charge carriers in silicon using a drift-diffusion model.
  Detailed electric field maps imported from TCAD simulations can be used to precisely model the drift behaviour of charge carriers within the silicon, bringing a new level of realism to Monte Carlo based simulations of particle detectors.

  This paper provides an overview of the framework and a selection of different simulation modules, and presents a comparison of simulation results with test beam data recorded with hybrid pixel detectors. Emphasis is placed on the performance of the framework itself, using a first-principles simulation of the detectors without addressing secondary ASIC-specific effects.

\end{abstract}


\end{frontmatter}

\section{Introduction}
\label{sec:introduction}

Detailed simulations of segmented silicon detectors are a crucial tool for understanding their performance and optimising their design.
Advanced tools for simulation such as TCAD exist, but are very demanding on computing time and often do not easily allow integration with other tools in order to facilitate a Monte Carlo approach, an essential method in high-energy physics given the stochastic nature of particle interactions.

While many custom packages such as AllPix~\cite{allpix}, KDetSim~\cite{kdetsim} or PixelAV~\cite{pixelav} implement standard drift-diffusion models of charge carriers, the software presented in this paper aims to provide a comprehensive, modular framework combining the simulation of material effects in the experimental setup (such as multiple scattering or nuclear interactions) with a detailed description of the motion of deposited charge carriers.

\apsq is a lightweight framework written in modern C++, with independent modules communicating through common objects passed between them using a message broker.
The structure of the framework is such that the development of new simulation algorithms can be performed with little knowledge of the underlying structure.
Encapsulation of external packages within a standalone module is an important part of simplifying the framework interface.

A comprehensive user manual has been prepared for the \apsq framework, describing its functionality and providing usage examples and answers to frequently asked questions.
A first version of this manual has been published~\cite{clicdp-apsq-manual}, while the most recent version should always be retrieved from the project website~\cite{apsq-website}.

This paper provides a brief overview of the \apsq framework (Section~\ref{sec:framework}) along with the simulation steps for a typical use case (Section~\ref{sec:simulation}).
A first validation with test beam data is presented in Section~\ref{sec:data}, and conclusions as well as an outlook to future developments are given in Section~\ref{sec:conclusion}.


\section{Framework Architecture}
\label{sec:framework}
\begin{figure}[tbp]
  \centering
  \includegraphics[width=.95\columnwidth]{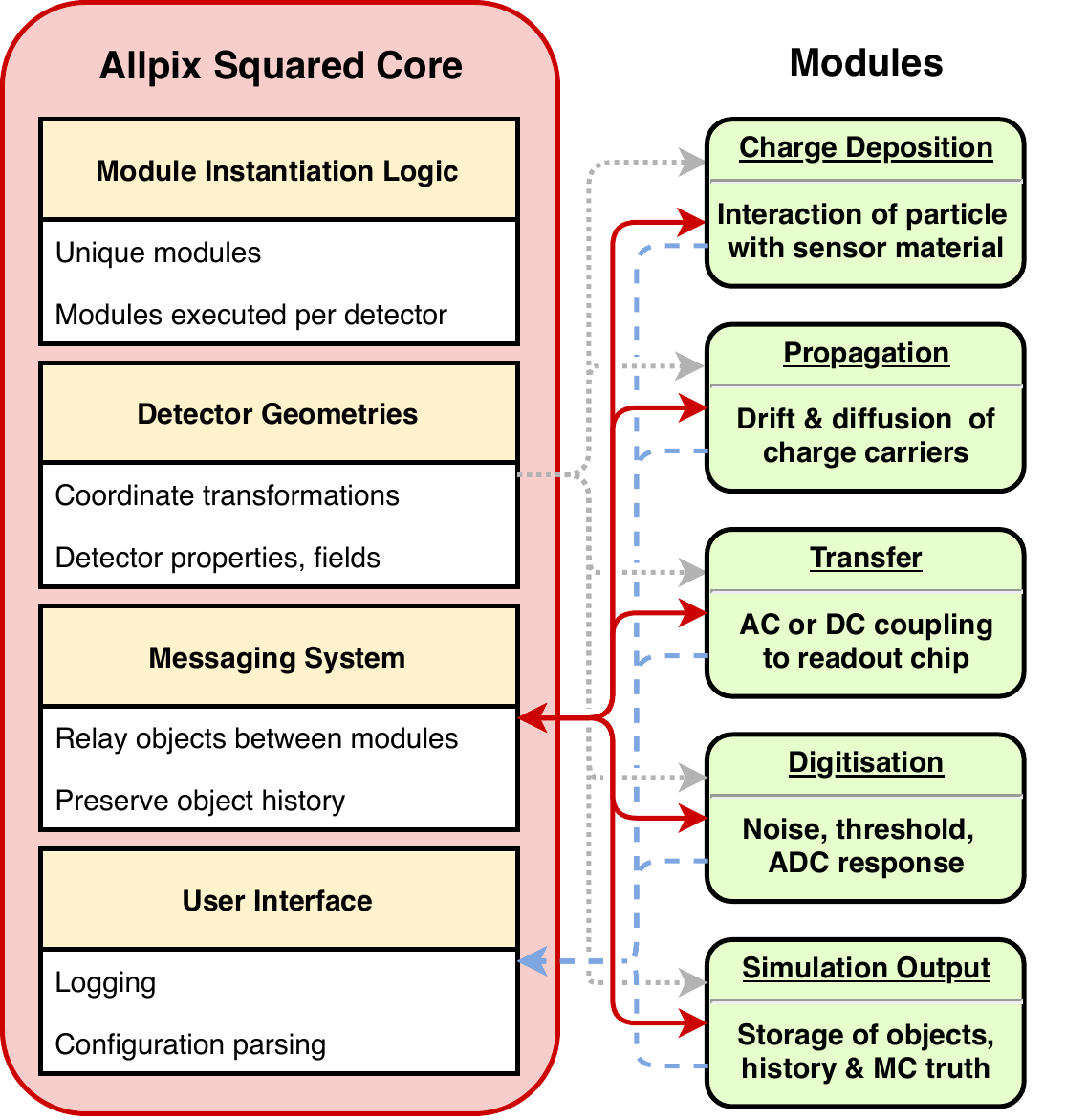}
  \caption[Structure of the \apsq framework]{Structure of the \apsq framework. The core of the framework (left) contains common utilities while individual modules (right) implement the simulation process. Grey dotted arrows indicate information about detector models, blue dashed arrows represent logging information, and red solid arrows constitute the bidirectional flow of messages.}
  \label{fig:framework}
\end{figure}

\apsq is built as a modular framework which separates central infrastructure components from the actual physics simulation implemented in individual modules.

\subsection{Framework Core}
\label{sec:framework_core}
The framework core provides four base components common to all modules in the simulation chain as indicated in Figure~\ref{fig:framework}: the module instantiation logic, the detector geometries, the messaging system, and the user interface.

\paragraph{Module instantiation logic}

The framework provides interfaces for two different types of modules.
\textit{Unique} modules perform tasks which require knowledge of the full detector setup, such as the simulation of a particle traversing all detectors and depositing energy; these modules are instantiated only once.
\textit{Detector} modules simulate processes within a single detector, such as the propagation of charge carriers in a single sensor; these modules only receive information about the detector in question and do not know about the global setup.
In this case, the simulation logic is simplified as the algorithm is only required to take care of a single detector, and allows more flexible configuration of the simulation chain as different modules can be used for different detectors.

The module instantiation logic of the framework deduces the number of module instantiations and their detector assignment directly from the configuration file provided by the user.
Module instances are created at startup of the framework and are used throughout the simulation process; statistics such as execution time are collected for each instance individually.

\paragraph{Detector geometries}

A flexible, parametrised geometry description is one of the key features of the framework since it facilitates changes to detector models without recompilation of the framework, and without changes to the actual program code.

The framework currently provides model descriptions for hybrid and monolithic pixel detectors, where hybrid pixel detectors consist of separate silicon blocks for the sensor and readout chip which are connected through bump bonds, and monolithic detectors consist of a single block of silicon.
Silicon strip sensors can be simulated in this manner by creating detector models with a single row of pixels and appropriate pitch.

In addition to the detector itself, an arbitrary number of support layers with different materials can be added to the model, with configurable dimensions and positioning.
These can encompass printed circuit boards, shielding, or protective covers around the detector, and each layer can optionally be fitted with a rectangular cut-out.

Models are described in standalone configuration files, a few of which are supplied with the framework.
Instances of these individual detectors are placed in the global reference frame of the simulation setup by specifying their type, position and orientation in the geometry configuration.
Individual parameters, such as the sensor thickness, can be modified for each placement of a model.

The framework calculates and provides access to all transformations between the global and the individual local coordinate systems, in addition to other detector properties such as electric fields.

\paragraph{Messaging system}

Information is exchanged between different modules through a messaging system.
Each module instance subscribes to a certain message type at startup, and all messages of the respective type, and potentially of the matching detector, are forwarded to the module during the simulation.
After processing the input data, modules can dispatch new messages to the system in order to relay results to other modules.
The messages hold a set of objects which form the result of the respective module, e.g.\ a set of charge carriers deposited at discrete positions in the active sensor material.

An important feature implemented in the \apsq framework is the object history:
for each object the full provenance is recorded, allowing reconstruction of the complete process for every single object.
For example, a charge pulse at the front-end amplifier would be connected to all propagated charge carriers which contributed to the reading.
The propagated charge carriers, in turn, would be related to the carriers created by the initial energy deposition and the particles simulated.
This feature thus allows direct relation of each pixel hit retrieved from the simulation to the initial particles and their exact position, i.e.\ to the Monte Carlo truth.

\paragraph{User interface}

\apsq provides a command line interface for interacting with the user.
All console output is redirected through a central logging facility which allows configuration of the verbosity, style and additional destinations of simulation logs, such as a text file.

\begin{figure}[tbp]
  \centering
  \setlength{\fboxsep}{0pt}%
  \fbox{\includegraphics[width=.95\columnwidth]{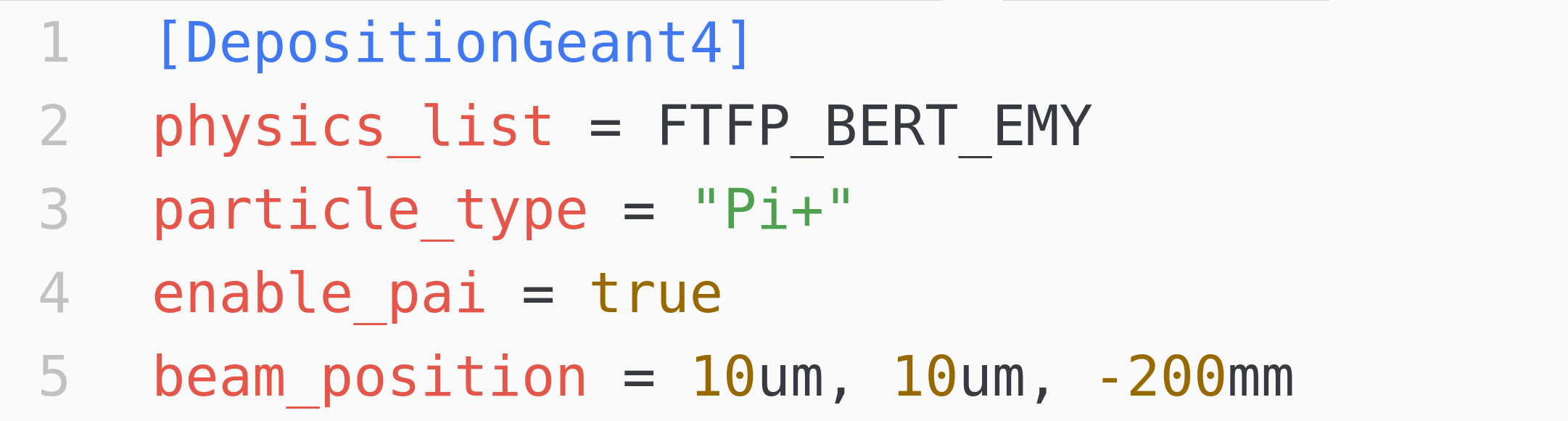}}
  \caption[]{Example configuration section with one header defining the "DepositionGeant4" module and four key-value pairs setting the Geant4 physics list, particle type, enabling of the PAI model, and specification of the position of the beam origin.}
  \label{fig:snippet}
\end{figure}

The framework itself, as well as the individual modules, is configured through minimalistic files using an intuitive syntax~\cite{tomlgit}.
Configuration files consist of section headers identifying modules, followed by a sequence of key-value pairs as shown in Figure~\ref{fig:snippet}.
Here, four parameters are set in lines 2--5 for the module specified by the header in line 1.

Configuration values support physical units, which are automatically converted into the units used internally by the framework.
This helps to avoid confusion or mistakes, by explicitly stating the physical unit for every input used in the simulation.
The value of the last key displayed in Figure~\ref{fig:snippet} is a vector with its components given in units of micrometres and millimetres, all of which will be automatically converted to the same unit for use inside the framework.

\apsq uses two configuration files to set up and execute the simulation.
The main configuration file contains one section per module of the simulation and defines both the framework parameters as well as the configuration of each individual module, while the detector configuration file describes the position of all detectors in the simulation setup as outlined above.
Several examples of simulation configurations for different use cases are provided in the framework's software repository~\cite{apsq-repo}, alongside a description of the setup and chosen parameters.

A more detailed description of the core framework functionality and its different options is provided in the user manual~\cite{clicdp-apsq-manual}.

\subsection{Software Development}
The development of \apsq follows best practices for software development by using the C++14 language standard~\cite{iso-cpp14}, adapting an agile development model, requiring strict format compliance and by enforcing a rigorous testing scheme.

The framework uses smart pointers to manage memory, \emph{auto} specifiers to counter type repetition, \emph{move} semantics for efficient transfer of resources and \emph{lambdas} for concise function implementations.
Furthermore, C++ templates are used throughout the framework to avoid code duplication and exceptions are used to significantly simplify error handling.
The code base is well documented and a full class reference is automatically generated from the source code using \texttt{doxygen}~\cite{doxygen} and provided on the website~\cite{apsq-website}.

Quality and compatibility of the \apsq framework is ensured by a continuous integration which builds and tests the software on all supported platforms.
It is integrated with the software repository and tests every new code submission to ensure the altered code can be compiled, that its formatting and code style adhere to the rules, and that all tests pass.
The test suite comprises three different types of tests.
Framework functionality tests aim at reproducing basic features such as correct parsing of configuration keys or resolution of module instantiations; module tests check the functionality of simulation modules and thus protect the framework against accidental changes affecting the physics simulation; and performance tests monitor the simulation speed.


\section{Simulation Flow}
\label{sec:simulation}
\begin{figure*}[tp]
  \center
  \resizebox{\textwidth}{!}{\includegraphics[width=\textwidth]{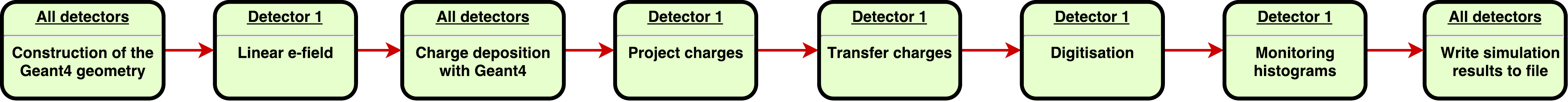}}
  \caption{Simple setup of an \apsq simulation chain with a single detector, where every block represents a single module instantiation. The modules are executed in the order they appear in the configuration file. First, additional geometry is constructed and an electric field is calculated for the detector. Then, charge carriers are deposited, propagated and collected at the implants. Finally, the signal is digitised and the simulation result is stored.}
  \label{fig:simulation:simple}
\end{figure*}

\begin{figure*}[tp]
  \center
  \resizebox{\textwidth}{!}{\includegraphics[width=\textwidth]{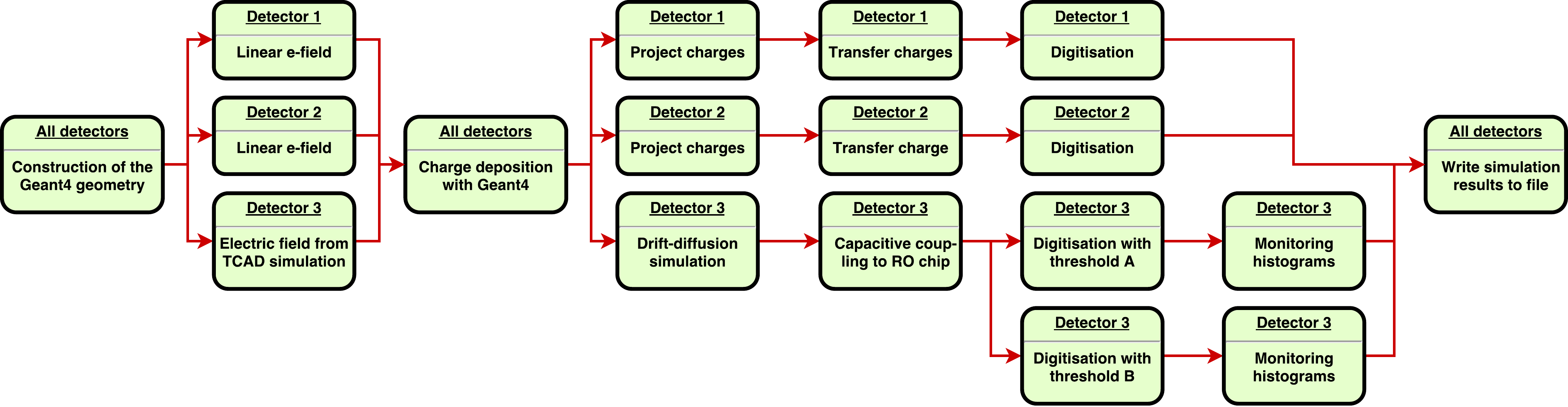}}
  \caption{Setup of an \apsq simulation chain with three detectors. Here, detectors 1 and 2 are treated as reference devices with a relatively simple simulation flow, while detector 3 is the device under test with more complex simulation modules. It uses a TCAD-generated electric field, the drift-diffusion propagation method and is capacitively coupled to its front-end electronics. The digitisation is performed with two different thresholds for comparison, where both datasets can be stored using different labels.}
  \label{fig:simulation:complex}
\end{figure*}

Every \apsq simulation consists of a set of modules and their specific parameters, executed sequentially for every simulated event.
While the main functionality of each module is carried out on an event-by-event basis, modules are given the option to run additional \emph{initialisation} code before the start of the simulation chain.
This can, for example, be used to process additional configuration information, such as the electric field generation described below.
After the last event has successfully been simulated, the \emph{finalisation} function of each module is called to allow for clean-up and to produce histograms.

While the modular approach of \apsq allows a wide variety of possible simulations to be carried out, a typical simulation involving a silicon detector and source of interacting particles will contain the steps described in the following paragraphs.
This example only represents a simple use case similar to the one outlined in Figure~\ref{fig:simulation:simple}, highlighting a few of the features currently implemented.
With little effort however, more complex simulation setups such as the sketch shown in Figure~\ref{fig:simulation:complex} can be produced, featuring different modules and configurations for multiple detectors.

\subsection{Generating Additional Geometry Information}

\begin{figure}[tbp]
  \centering
  \begin{overpic}[width=.9\columnwidth]{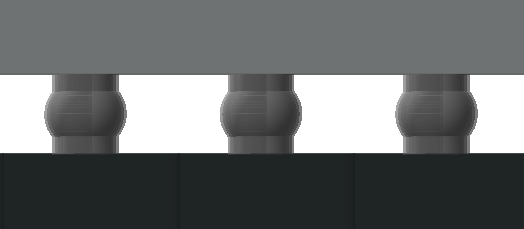}
      \put (5,35) {Sensor}
      \put (5,5) {\textcolor{white}{Readout ASIC}}
  \end{overpic}
  \caption{Visualisation of three bump bonds between the sensor and readout ASIC of a hybrid detector, approximated by joined volumes of a sphere and a cylindrical column.}
  \label{fig:bumps}
\end{figure}

While all transformations from local to global coordinates are calculated by the framework core, some modules may require additional information beyond the detector models, such as materials.
For this purpose, modules are allowed to store additional objects alongside the detectors.
One example is the \emph{GeometryBuilderGeant4} module, which creates a geometry description compatible with Geant4~\cite{geant4,geant4-2,geant4-3} from the internal detector models by using native Geant4 object types.
The model's configuration keys are queried in order to perform pre-defined sequences which yield, for example, typical hybrid or monolithic pixel detector layouts.

The separation of this step into an individual module instead of direct implementation in the framework core minimises external dependencies and allows operation of the framework without the respective dependency, e.g. for further processing of previously simulated data.

In the Geant4 model, the hybrid pixel detector model described earlier consists of a silicon sensor and readout ASIC, and of a connection via an approximation of bump-bonds.
These are implemented as the joined volumes of a sphere and cylindrical column, the parameters of which are configurable.
A zoom of such a structure is shown in Figure~\ref{fig:bumps}.
Monolithic pixel detectors contain both sensor and readout circuitry in the same silicon wafer, and are thus implemented as a single block of silicon.

\subsection{Visualisation of the Setup}
\label{sec:simulation:viz}

The ability to perform a 3D visualisation of the detector setup is a powerful tool to verify the configuration and placement of detectors, in particular where several detectors are simulated together.
\apsq provides the \emph{VisualizationGeant4} module which uses the previously described Geant4-compatible detector models and acts as an interface to the internal viewers provided by the Geant4 framework.
The visualisation includes all detectors placed in the setup as well as primary and secondary particles simulated by Geant4.
An example of such an event display is shown later in Section~\ref{sec:data}.

\subsection{Electric Fields}

As in a physical detector, \apsq does not consider the electric field in the sensor volume as part of the detector geometry.
Instead, electric fields are added by a dedicated module, \emph{ElectricFieldReader}, which represents the equivalent of switching on the high voltage power supply in the laboratory or test beam measurement.
This not only simplifies the detector configuration, but facilitates the application of different electric fields or bias voltages to the same type of detector.

Currently, two types of electric field profiles can be applied to detectors: a linear electric field or an imported field map from TCAD.
In the case of a linear electric field, the depletion voltage or depletion depth of the sensor can be specified along with the applied bias voltage, and a field will be calculated which varies linearly inside the depleted region of the sensor.

The more detailed approach involves the import of an electrostatic field simulation from packages such as TCAD.
Here, \apsq comes with an independent converter tool to convert the electric field distribution from the adaptive mesh produced by the TCAD simulation of a single pixel cell to a uniformly-spaced grid.
By this conversion, the computation time for an interpolation between different grid points is greatly reduced.

The resulting electric field is replicated for all pixels across the matrix, and allows detailed studies of charge collection to be carried out.
As the electric field is accessed via the detector model for a given 3D point irrespective of the electric field applied to the detector, the field lookup is transparent to the level of detail (or method used) for the supplied field.

\subsection{Deposition of Charge Carriers}

While different modules for the initial deposition of charge carriers in the sensor can be envisioned, currently only one such module is implemented in \apsq: \emph{DepositionGeant4}.
This module acts as an interface to Geant4, which carries out the generation and propagation of particles throughout the volume surrounding the setup.
In addition to configuring Geant4's particle properties according to the user-provided configuration, the maximum stepping size for particle transport can be configured.
A range threshold for the production of $\gamma$, e$^-$ and e$^+$ is necessary to avoid infrared divergence.
The module automatically calculates an appropriate range cut based on the minimal feature size of a single pixel cell in the simulation, rather than relying on the default value provided by Geant4.
By default, a fifth of the minimal feature size of any detector in the setup is used.
In this way, the description of effects from secondary particles on the cluster size is significantly improved.

Any of the available physics lists from Geant4 can be used to describe the particle interactions.
Optionally, the Photoabsorption Ionisation model (PAI) can be activated to improve the modelling of very thin sensors~\cite{pai}.

All energy deposits that are produced within sensitive volumes are converted to charge carrier deposits and dispatched via the framework messaging system for further use.
In addition, information about both the primary particle trajectory and any secondary particles created such as delta electrons is stored and linked to the charge carriers.

\subsection{Propagation of Charge Carriers}
\label{sec:propagation}

The description of the propagation of charge carriers through the sensitive volume is one of the key components of any semiconductor detector simulation.
Currently two different methods, with varying complexity, are implemented in \apsq.

The most simple -- and thus fastest -- of these is the \emph{ProjectionPropagation} module.
This module calculates the total drift time for each set of charge carriers using an analytical approximation for the integral of the mobility in a linear electric field.
A randomised lateral diffusion is calculated from a two-dimensional Gaussian distribution according to the Einstein formulation, and the charge carriers are placed at the surface of the sensor.
For thick, planar silicon sensors, this pragmatic approach produces sufficiently precise results, especially if the respective detectors are only used for reference.

A more detailed algorithm is implemented in the so-called \textit{GenericPropagation} module, which uses the Jacoboni parametrisation of mobility~\cite{JACOBONI197777} to describe carrier motion throughout the sensor.
Groups of charge carriers of user-defined size are stepped through the sensor, taking into account the electric field by means of a Runge-Kutta integration~\cite{fehlberg} and the calculation of the mobility at each position.
Random diffusion is added to every step of the drift motion.

As the integration of the electric field is performed using a request for field vectors in 3D space, the method used to evaluate the electric field is transparent to the propagation module.
Where a field map from TCAD is used this feature becomes a very powerful way to observe charge flow along field lines within the sensor.
This can be of particular interest where complicated field profiles arise such as for depleted CMOS devices with strongly non-linear fields and potentially undepleted regions.

At present, the propagation of groups of charges throughout the sensor is performed in isolation, without the calculation of transient currents induced on the charge collection implants.
For many applications, this abstraction is sufficient for a suitable description of the detector performance.
The implementation of induced current simulation, with correct treatment of the Ramo weighting field~\cite{shockley,ramo}, is currently ongoing as briefly discussed in Section~\ref{sec:conclusion}.

\subsection{Transfer from Sensor to Readout Chip}

\apsq introduces an additional step between the propagation of charge carriers and their digitisation in the front-end electronics, namely the transfer of the signal from sensor to readout chip.
This step adds additional flexibility to the simulation chain for detectors with different interconnect technologies.

In monolithic sensors, or in hybrid detectors where the sensor is DC-coupled to the readout electronics by bump bonds, the signal is formed by the \emph{SimpleTransfer} module by simply grouping together all charge carriers located within the pixel boundary, within a given depth of the pixel implant.

For hybrid pixel detectors with a capacitively coupled sensor, stray capacitances to the neighbouring pixels may induce additional signals and thus pixel hits.
In this scenario, the \emph{CapacitiveTransfer} module can be used to simulate cross-coupling between different pixels in order to emulate the behaviour of the assembly.

\subsection{Digitisation of the Signal}

Using the charge transferred to a given pixel, a generic digitisation module is used to simulate the response of the front-end pixel electronics.
This is not intended to represent all available front-end ASICs, but contains many configurable parameters which allows a variety of chip architectures to be simulated.

The input charge can be combined with electronic noise before comparing with a threshold with optional dispersion.
A conversion factor from electrons to ADC units can be used to simulate preamplifier gain and to map the charge onto a counter of customisable precision.
In combination with a fixed offset, this can be used to simulate the response of a typical Time-over-Threshold (ToT) digitisation.

\subsection{Storing the Simulation Result}

The native data format of \apsq is a collection of ROOT~\cite{root} \texttt{trees}, each tree storing information for a specific detector and message type.
These trees can be produced and written to file by the module \emph{ROOTObjectWriter}, which in addition also stores a full copy of the framework and module configuration as well as the detector setup used in the respective simulation.
This means that a full simulation can be reproduced from the information contained in a single data file, and one can refer to the stored configuration in order to verify specific settings for the simulation under investigation.

The information stored in the trees can be controlled, and certain information can be explicitly included or excluded from storage, such as the original position of deposited energy, or the entry and exit points of primary particles in the sensors.

A powerful feature of this data format is its ability to be replayed.
The \emph{ROOTObjectReader} module can be used to read stored simulation data from such a file and to dispatch every message to the framework.
This allows the saving of intermediate stages of the simulation to disk, e.g.\ post charge collection, and to restart from this stage several times in order to execute the final step of digitisation with variable settings without the need to recompute the full simulation chain.

In addition to this native format, \apsq comes with a variety of output modules to support several currently-used reconstruction and analysis frameworks, such as EUTelescope~\cite{EUDET-2010-12}, Proteus~\cite{judith,proteus} or Corryvreckan~\cite{corry}; the last of which has been used to obtain the results presented in the following section.


\section{Comparison with Testbeam Data}
\label{sec:data}
Any new simulation software has to be validated in order to demonstrate its ability to accurately describe data.
For this purpose, the data presented in~\cite{AlipourTehrani} are compared to simulations performed with the \apsq framework.
This comparison includes the simulation of a beam telescope and the corresponding reconstruction of particle tracks, as well as a detailed description of the properties of one of the detectors in the setup, the so-called device under test (DUT).

While the emphasis of this paper is to showcase the performance of the simulation framework, rather than to achieve the best possible agreement between data and simulation by detailed tuning, the current \apsq modules are implemented in a first-principles approach which is expected to accurately reproduce data.
This requires applying only a few crucial parameters taken from the data analysis, without taking into account additional secondary (and detector-specific) effects such as cross-coupling in the front-end chip, noisy pixels or per-pixel gain variations, which may affect the performance of the detectors.

\begin{table}[tbp]
  \centering
  \caption{Operating parameters for the devices under test and the telescope planes, comparing the values from data with those used in the simulation. Values for data are taken from~\cite{AlipourTehrani}.}
  \label{tab:parameters}
  \begin{tabular}{lrr}
    \toprule
    Parameter         & Data & Simulation \\
    \midrule
    Temperature       & \SI{293}{\kelvin} & \SI{293}{\kelvin} \\
    Electronic noise  & \SI{\sim 80}{e^-} & \SI{100}{e^-} \\
    Gain factor       & \multicolumn{1}{c}{---} & \num{1.03} \\
    \midrule
    \multicolumn{3}{c}{\SI{50}{\micro\metre} DUT} \\
    \midrule
    Depletion voltage &  \SI{-7}{\V} &  \SI{-7}{\V} \\
    Bias voltage      & \SI{-15}{\V} & \SI{-15}{\V} \\
    Threshold         & \SI{506\pm31}{e^-} &  \SI{500\pm30}{e^-}\\
    ADC smearing      & \multicolumn{1}{c}{---} & \SI{350}{e^-}\\
    \midrule
    \multicolumn{3}{c}{\SI{100}{\micro\metre} DUT} \\
    \midrule
    Depletion voltage & \SI{-12}{\V} & \SI{-12}{\V} \\
    Bias voltage      & \SI{-20}{\V} & \SI{-20}{\V} \\
    Threshold         & \SI{537\pm33}{e^-} & \SI{520\pm35}{e^-} \\
    ADC smearing      & \multicolumn{1}{c}{---} & \SI{600}{e^-}\\
    \midrule
    \multicolumn{3}{c}{Telescope planes} \\
    \midrule
    Depletion voltage & \SI{+30}{\V} & \SI{+30}{\V} \\
    Bias voltage      & \SI{+50}{\V} & \SI{+50}{\V} \\
    Threshold         & \SI{\sim 1000}{e^-} & \SI{1000}{e^-} \\
    \bottomrule
  \end{tabular}
\end{table}

The only input to this simulation has been the beam properties and the Geant4 physics list, the geometry of the experimental setup, and the operating parameters listed in Table~\ref{tab:parameters}.
Since no per-pixel gain variation and corresponding calibration has been simulated, two parameters of the charge ADC simulation have been tuned in order to match the charge-calibrated results obtained from data.
The first parameter is a Gaussian smearing applied to the signal after digitisation, which represents a convolution of the reduced precision resulting from the ToT measurement of the readout ASIC, and the per-pixel gain and threshold variations.
The different values used for the DUTs are listed in Table~\ref{tab:parameters}.
The second adjusted parameter is a gain factor for the pixel charge, and a value of \num{1.03} has been used to allow for a comparison of the distribution shapes.
This correction is well within the uncertainty of the charge calibration.

\subsection{Experimental Setup}

The experimental setup consists of a telescope constructed of six planes arranged symmetrically around the DUT in the centre, with three planes in the upstream arm and three in the downstream arm of the telescope.
The individual telescope planes consist of Timepix3 ASICs~\cite{timepix3} with a thickness of \SI{700}{\micro\metre}, bump bonded to \SI{300}{\micro\metre} thick planar ${\textit{p-in-n}}$ silicon sensors.
The detectors are glued to support PCBs which hold electronic components required for the operation of the ASIC.

Each of these planes is rotated by 9\textdegree\ from perpendicular incidence in both the ${\textit{xz-}}$ and ${\textit{yz-}}$planes, with the ${\textit{z-}}$axis given by the beam direction.
In addition, the upstream arm is rotated by 180\textdegree\ around the ${\textit{y-}}$axis such that all sensors face the DUT and the distance between upstream and downstream arms is minimised.

Data from two different DUTs will be considered, both consisting of Timepix3 ASICs bump-bonded to ${\textit{n-in-p}}$ silicon sensors with thicknesses of \SI{50}{\micro\metre} and \SI{100}{\micro\metre} respectively.
Table~\ref{tab:parameters} lists the relevant operating parameters of the DUTs as well as the telescope planes.

The tracks reconstructed from data have been selected for comparison with the DUT based on their $\chi^2$ values, requiring $\chi^2/\textrm{ndf} < 100$, and a spatial cut of \SI{50}{\micro\metre} in both the ${\textit{x-}}$ and ${\textit{y-}}$directions was applied in order to associate them to the corresponding DUT cluster.
Cluster positions on the DUT were calculated as the charge-weighted centre-of-gravity using an $\eta$-correction method~\cite{Belau1983253}.
The experimental setup, the reconstruction and the analysis results are described in greater detail in~\cite{AlipourTehrani}.

\subsection{Simulated Setup}

\begin{figure}[tbp]
  \centering
  \includegraphics[width=\columnwidth]{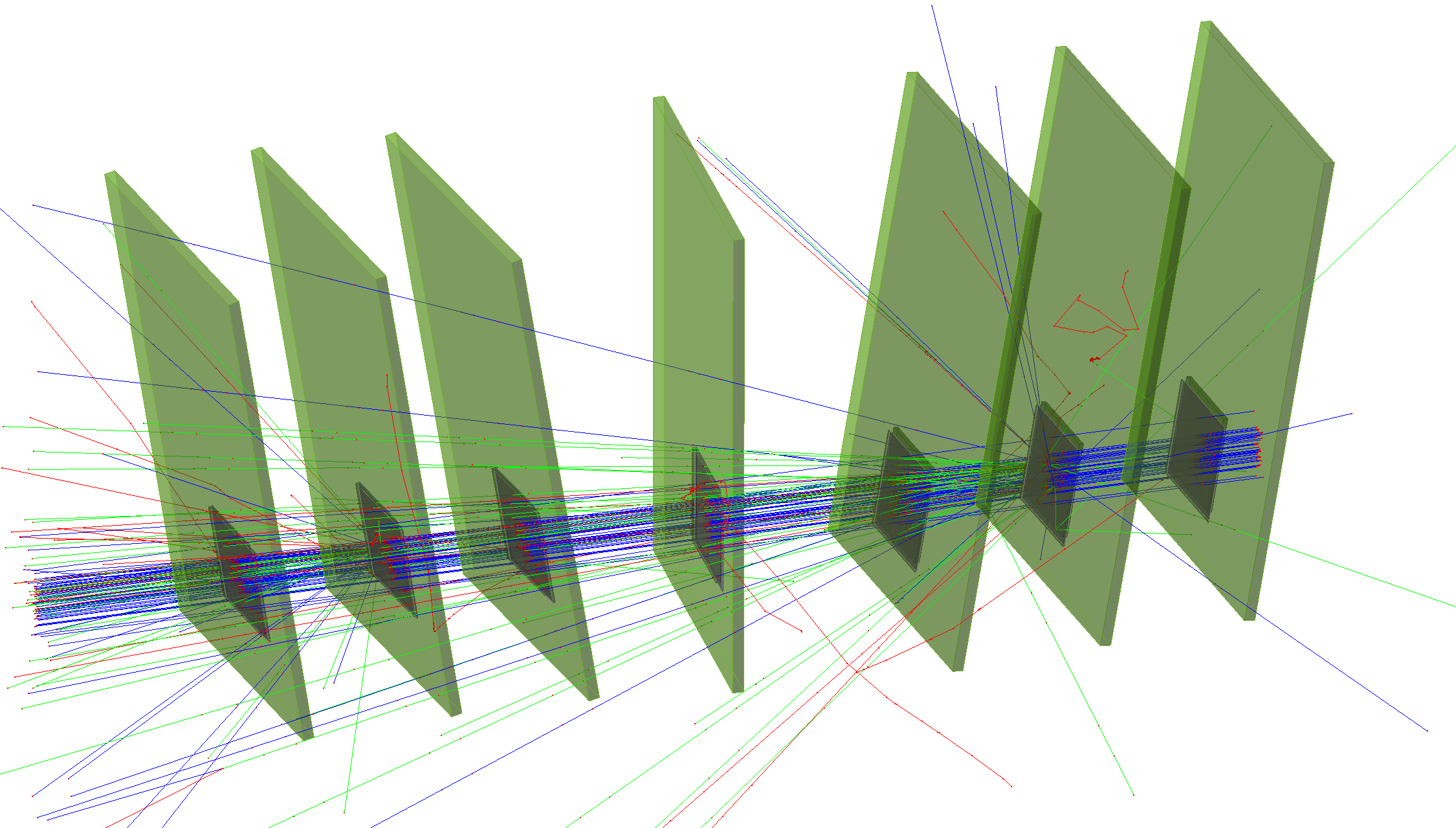}
  \caption{Visualisation of the simulated detector setup with six rotated telescope planes and one DUT in the centre, oriented perpendicular to the beam incident from the right. The colored lines represent the primary and secondary particles propagated through the setup.}
  \label{fig:telescope}
\end{figure}

The simulated detector geometry resembles the experimental setup and is shown in Figure~\ref{fig:telescope} using the \emph{VisualizationGeant4} module described in Section~\ref{sec:simulation:viz}.
The telescope planes are placed on an epoxy-based PCB with a thickness of \SI{1.76}{\milli\metre} and the lead-tin bump bonds between ASIC and sensor are modelled as shown in Figure~\ref{fig:bumps}, with a total height of \SI{20}{\micro\metre} and a sphere radius of \SI{9}{\micro\metre}.
This translates to an average material budget of about $\SI{2}{\percent}\,X_0$ per telescope plane.

The planes are positioned as described above, with an additional misalignment added to their exact positions and orientations by random sampling of Gaussian distributions with widths of \SI{1}{\milli\metre} and 0.2\textdegree~respectively.
This measure avoids pixel-perfect alignment in the simulation and thus reduces quantisation effects and anomalous track fit results.
For both DUTs, the \textit{GenericPropagation} module described in Section~\ref{sec:propagation} is used and the digitisation parameters are taken from data, with the values listed in Table~\ref{tab:parameters}.

The simulated particle beam consists of \SI{120}{\GeV}~\Ppiplus particles with a radial Gaussian width of \SI{2}{\milli\metre}; only a single particle is simulated per event, resembling the beam used for the measurements.
The Geant4 physics list \texttt{FTFP\_BERT\_EMY}, enabling the \emph{EM Standard Option 3}, is chosen to describe the interaction with the sensor material, and the PAI model is enabled to improve the description of energy deposition in very thin silicon sensors.
The simulation results are stored as ROOT \texttt{trees} for reconstruction.


\subsection{Simulation Results and Comparison with Data}

Cluster positions and particle tracks of the simulation were reconstructed using the \corry framework.
Errors for the track fit were taken from data, using the previously measured hit resolution of the telescope planes.
Track association and ${\chi}^2$ criteria equivalent to those used in the data analysis were applied for the track quality and DUT cluster selection, and a similar ${\eta}$-correction procedure was applied to the cluster position.
The random shifts and rotations applied to the initial position of the detectors were reproduced correctly by the alignment procedure.

\paragraph{Telescope track reconstruction}

\begin{figure}[tbp]
  \centering
  \includegraphics[width=\columnwidth]{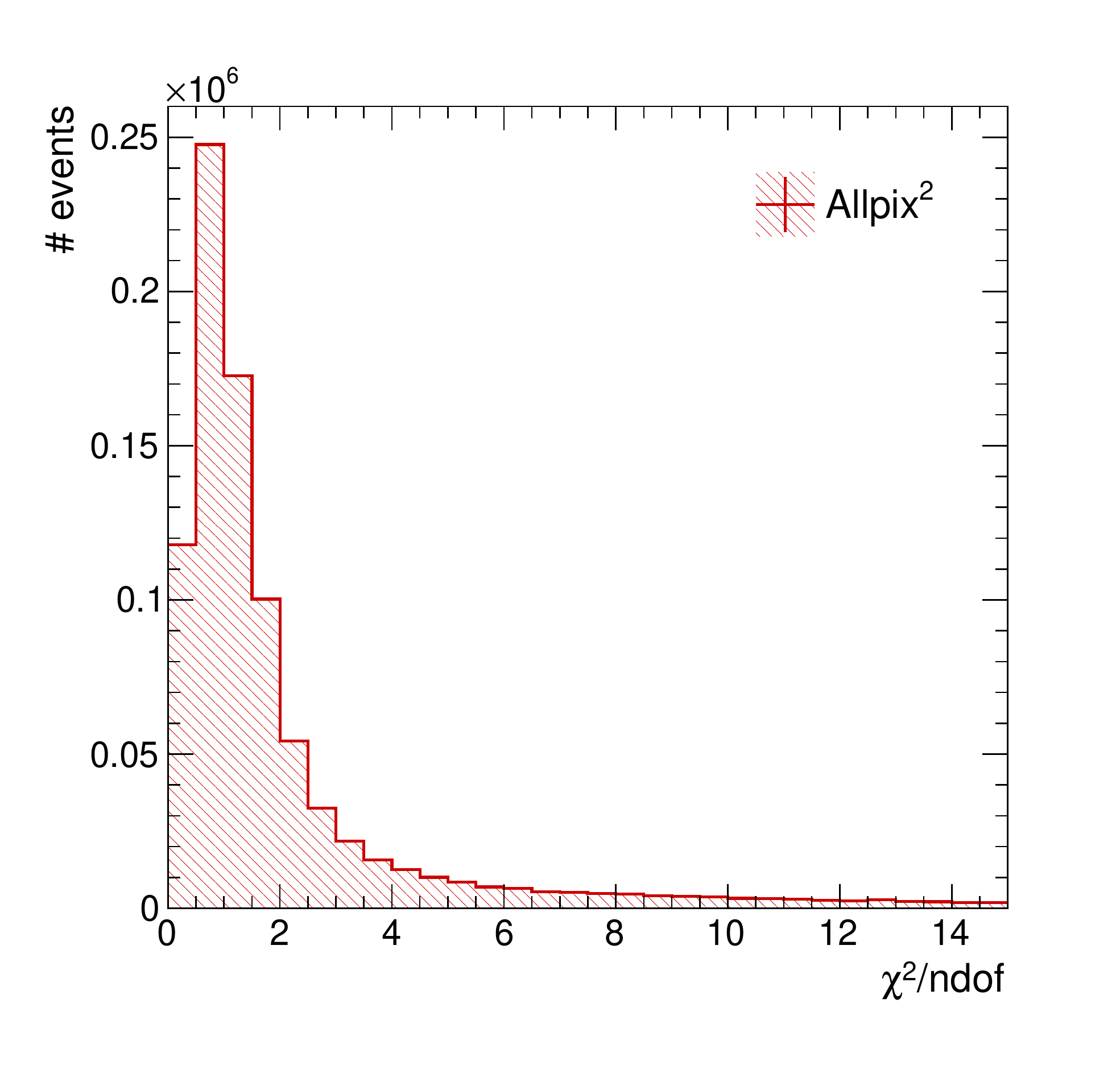}
  \caption{$\chi^2$ over number of degrees of freedom for straight particle tracks through the beam telescope reconstructed from simulation.}
  \label{fig:trackchi2}
\end{figure}

Particle tracks were fit using a straight line approximation via a linear regression.
Owing to the primary particle's high energy of \SI{120}{\GeV} this is expected to provide an adequate description of their trajectory.
The resulting distribution of the $\chi^2$ value over the number of degrees of freedom shown in Figure~\ref{fig:trackchi2} indicates an excellent goodness of fit with a most probable value of one.

\begin{figure}[tbp]
  \centering
  \includegraphics[width=\columnwidth]{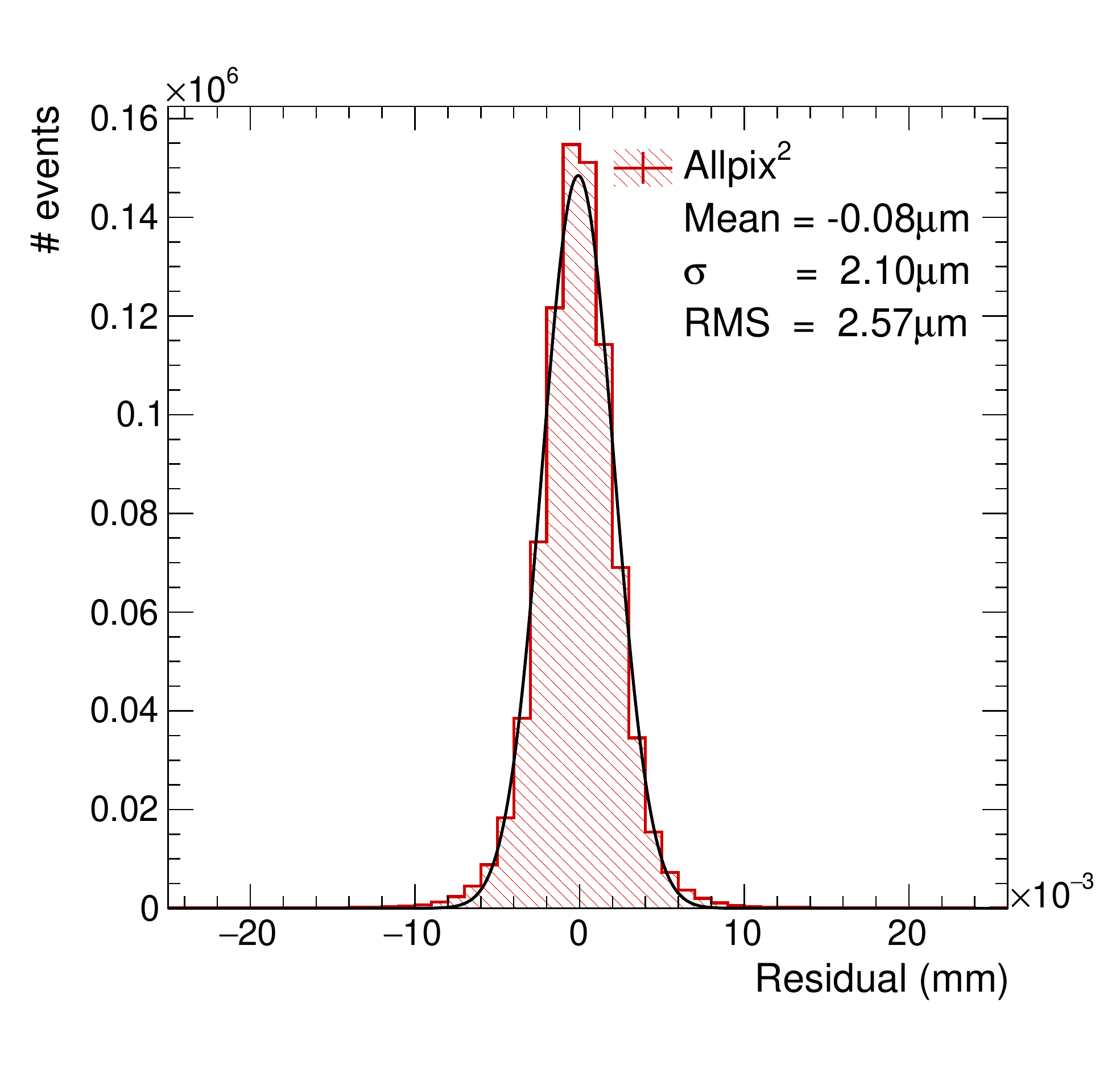}
  \caption{Telescope track residuals at the position of the DUT, calculated as the difference between the reconstructed track position and the known \emph{true} position of the simulated particle.}
  \label{fig:telres}
\end{figure}

Taking advantage of the known entry and exit points of the initial Monte Carlo particle traversing the DUT sensor, it is possible to extract the spatial precision of tracks reconstructed in the telescope by comparing the track intercept at the DUT with the known midpoint of the particle trajectory in the DUT sensor.
This residual is shown in Figure~\ref{fig:telres}, indicating a spatial track resolution at the position of the DUT of about \SI{2}{\micro\metre}, measured as the width of a Gaussian distribution fitted to the histogram.
This value is in agreement with the telescope residuals obtained from data~\cite{AlipourTehrani}.

\paragraph{DUT cluster size and charge collection}

\begin{figure}[tbp]
  \centering
  \includegraphics[width=\columnwidth]{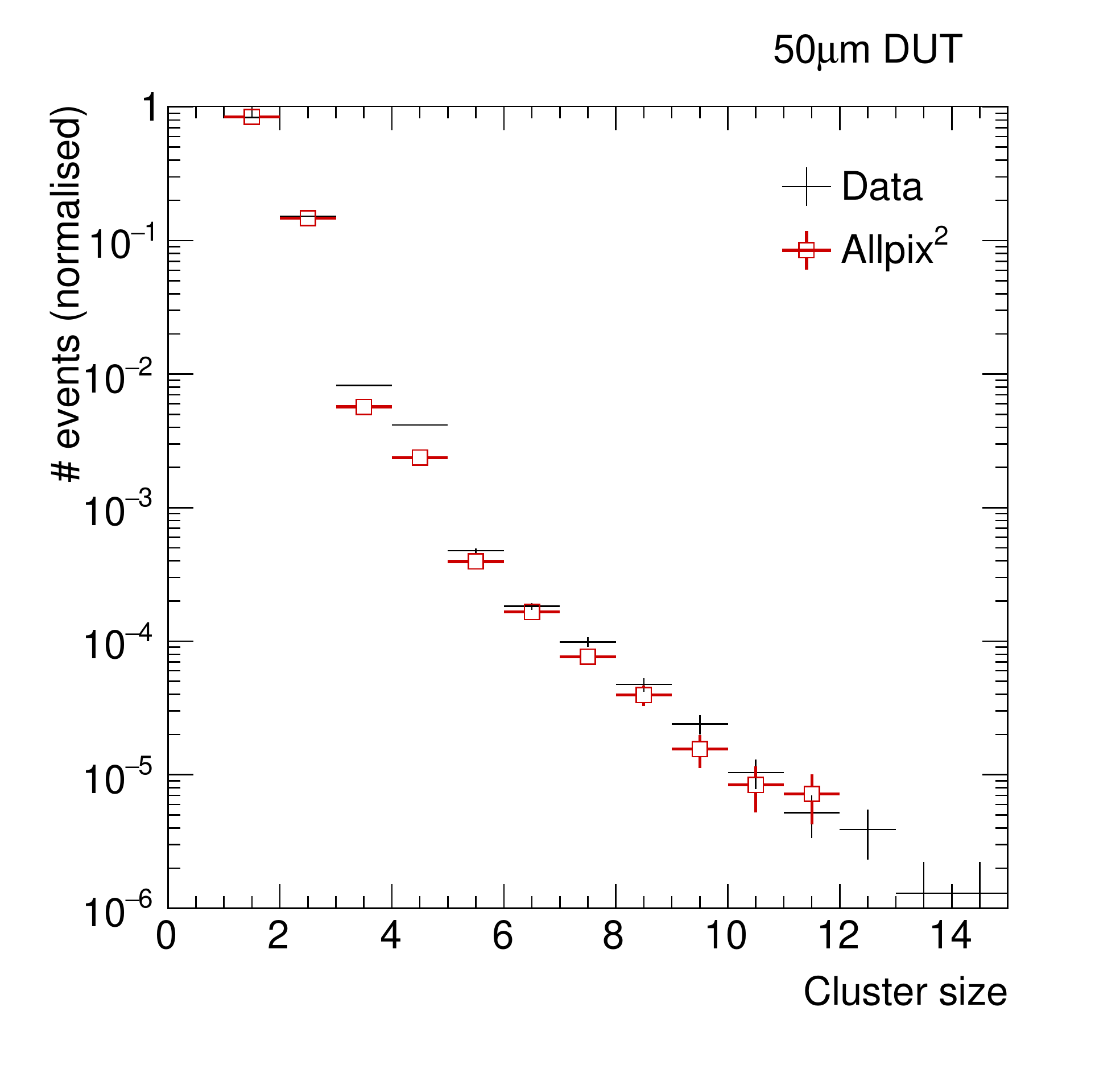}
  \caption{Comparison of the total cluster size in data and simulation for the \SI{50}{\micro\metre} thick DUT.}
  \label{fig:cluster:50}
\end{figure}

\begin{figure}[tbp]
  \centering
  \includegraphics[width=\columnwidth]{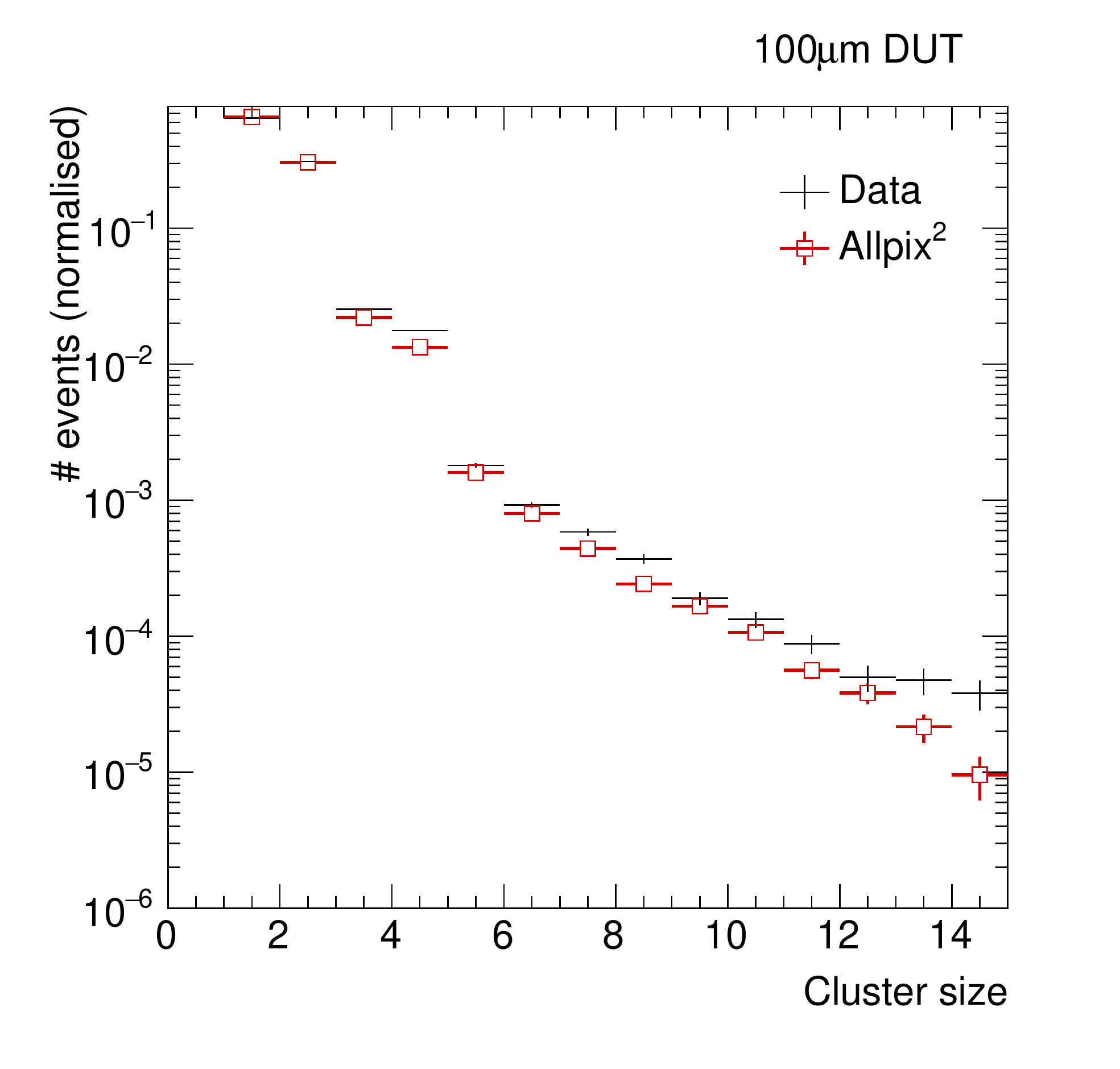}
  \caption{Comparison of the total cluster size in data and simulation for the \SI{100}{\micro\metre} thick DUT.}
  \label{fig:cluster:100}
\end{figure}

The distribution of the signal over several pixels and the resulting distribution of cluster sizes allows to gauge the charge propagation process as well as the effect of the threshold applied to the signal in the individual pixels.
Figures~\ref{fig:cluster:50} and~\ref{fig:cluster:100} demonstrate an excellent agreement between data and simulation for the spatial extent of the charge collected, indicating that the modelled drift and diffusion processes describe the data accurately.
The small difference observed for three- and four-pixel clusters can be attributed to variations in the actual threshold of the devices and could be reduced further by tuning the threshold values applied in simulation.
It should be noted that the agreement extends even to the tails with high cluster sizes, which are dominated by higher order effects such as delta electron emission.

\begin{figure}[tbp]
  \centering
  \includegraphics[width=\columnwidth]{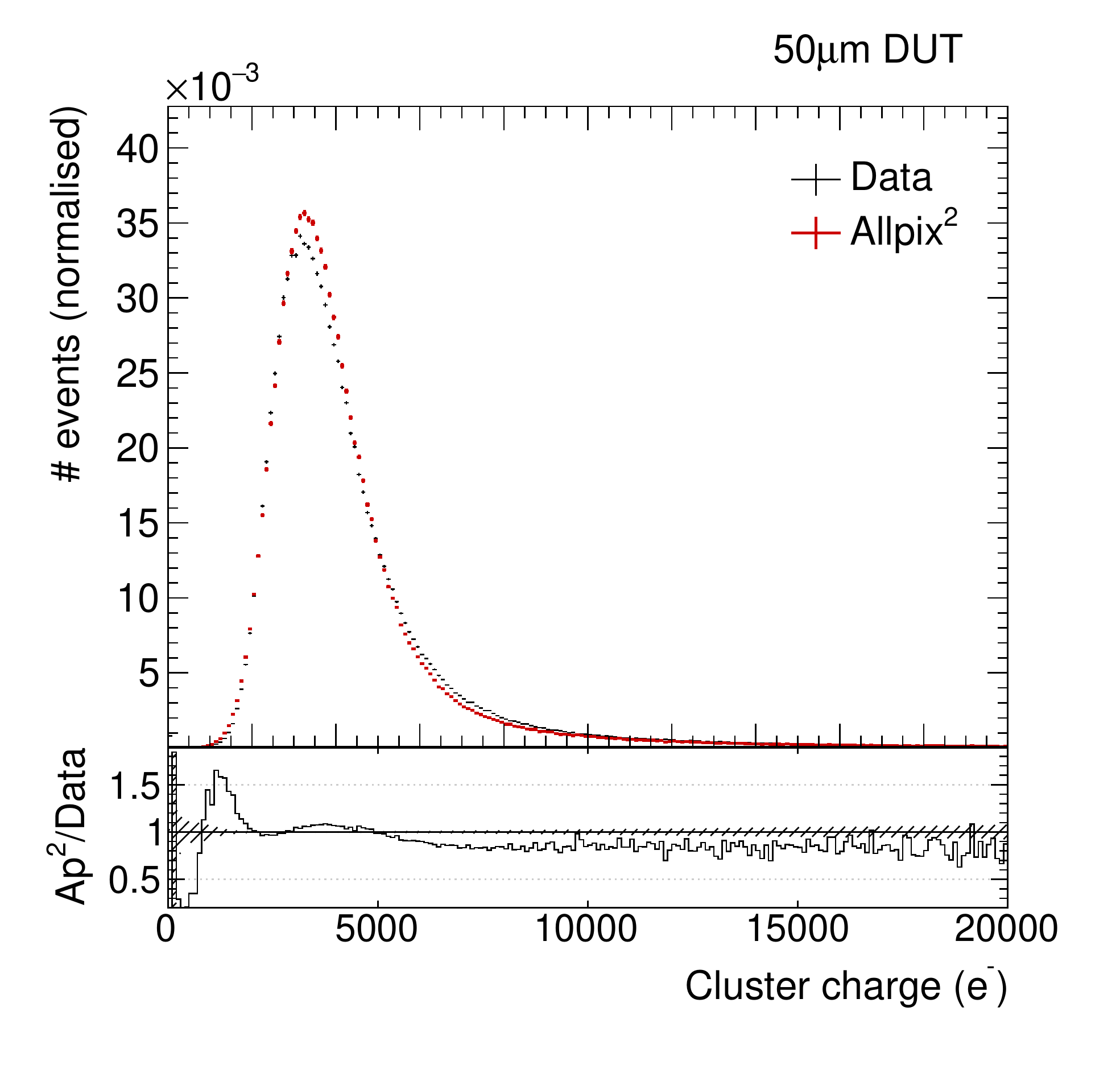}
  \caption{Cluster charge distribution for data and simulation of the \SI{50}{\micro\metre} thick DUT. The hatched band represents the statistical uncertainty on data.}
  \label{fig:landau:50}
\end{figure}

\begin{figure}[tbp]
  \centering
  \includegraphics[width=\columnwidth]{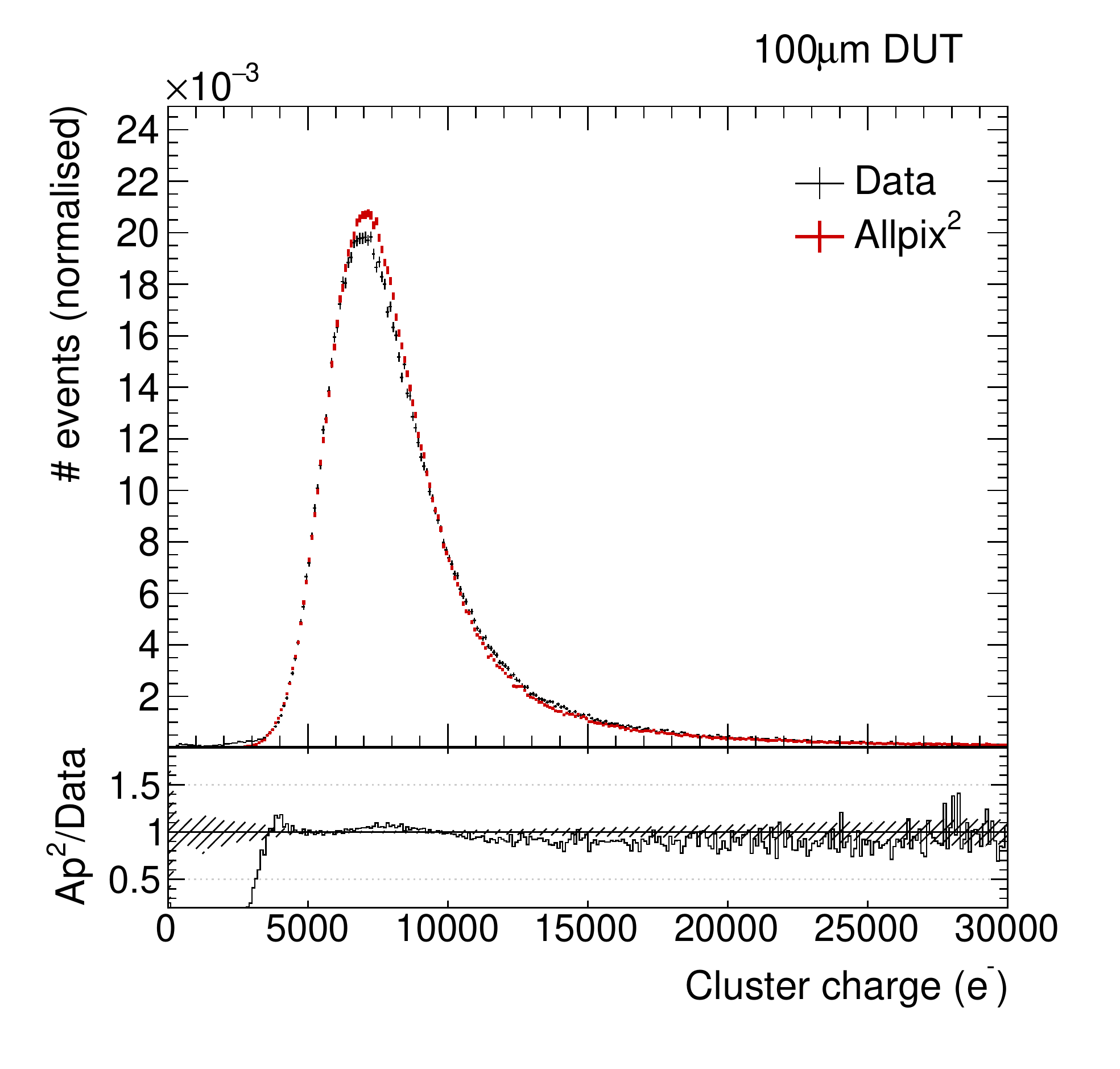}
  \caption{Cluster charge distribution for data and simulation of the \SI{100}{\micro\metre} thick DUT. The hatched band represents the statistical uncertainty on data.}
  \label{fig:landau:100}
\end{figure}

The total charge collected per cluster is shown in Figures~\ref{fig:landau:50} and~\ref{fig:landau:100} for the \SI{50}{\micro\metre} and \SI{100}{\micro\metre} thick sensors respectively.
The shape closely resembles the Landau distribution measured in data.
A further improvement of the agreement is expected with a more elaborate simulation of the per-pixel threshold and gain variations in the charge calibration of the ASIC.

\paragraph{Intrinsic resolution of the DUT}

\begin{figure}[tbp]
  \centering
  \includegraphics[width=\columnwidth]{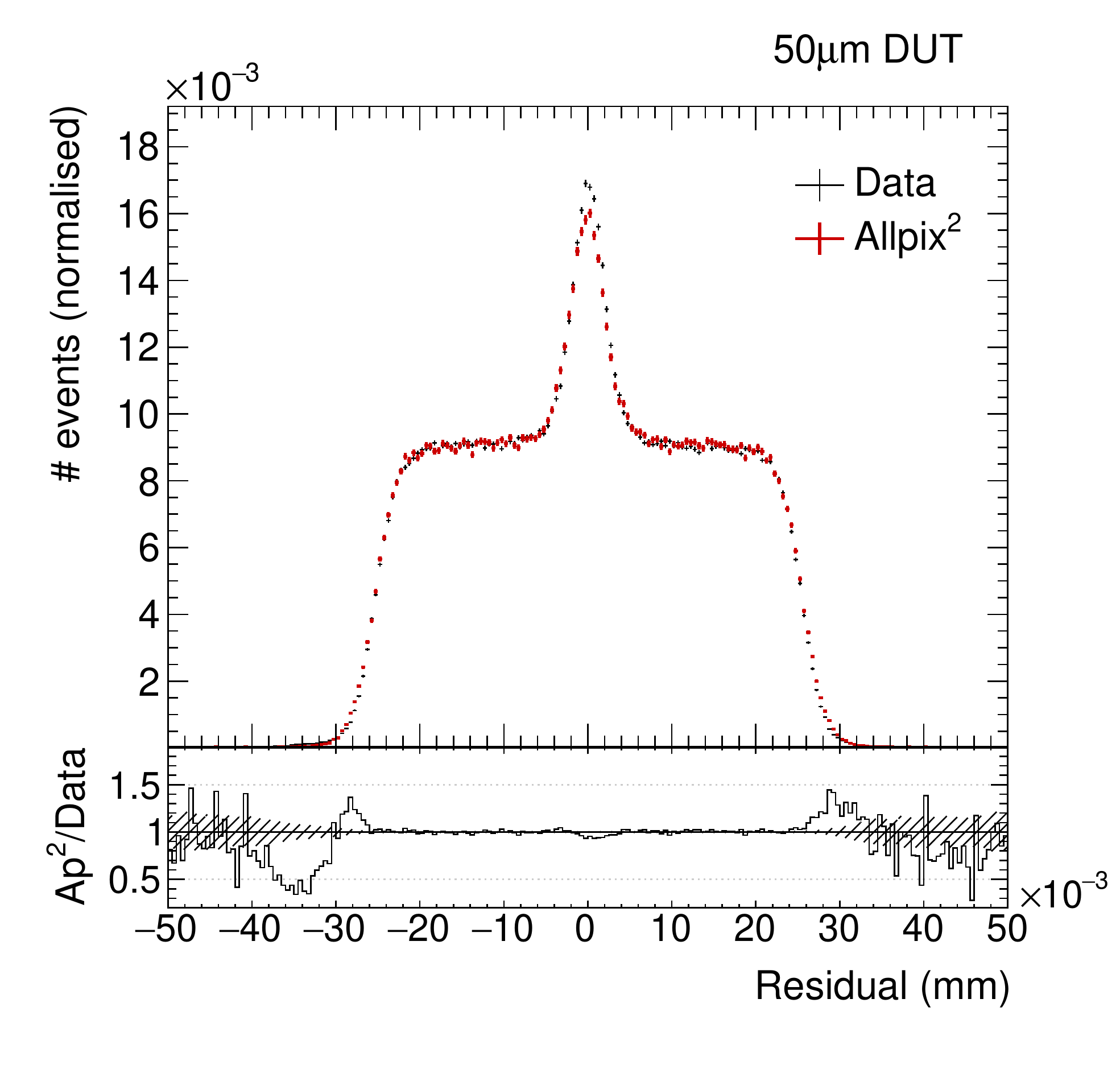}
  \caption{Track residual distribution for the \SI{50}{\micro\metre} thick sensor for data and simulation. Calculated as the difference between the track position and the reconstructed position of the particle in the DUT. The hatched band represents the statistical uncertainty on data.}
  \label{fig:residuals:50}
\end{figure}

\begin{figure}[tbp]
  \centering
  \includegraphics[width=\columnwidth]{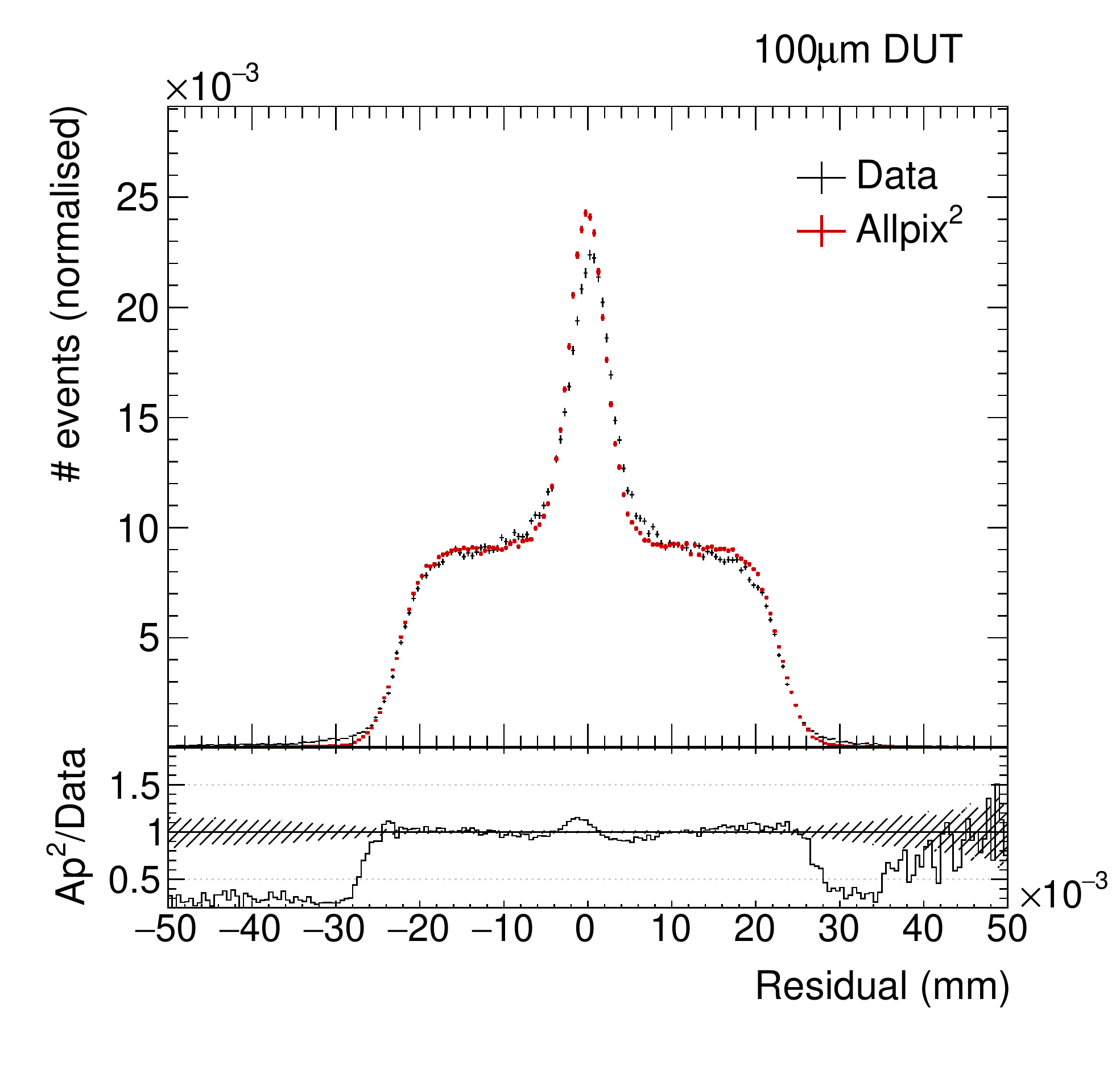}
  \caption{Track residual distribution for the \SI{100}{\micro\metre} thick sensor for data and simulation. Calculated as the difference between the track position and the reconstructed position of the particle in the DUT. The hatched band represents the statistical uncertainty on data.}
  \label{fig:residuals:100}
\end{figure}

The intrinsic resolution of the DUTs can be measured by calculating the unbiased track residual at the DUT as shown in Figures~\ref{fig:residuals:50}~and~\ref{fig:residuals:100} for the two DUTs, and by quadratically subtracting the track resolution from the measured residual width.

While the two curves match for the \SI{50}{\micro\metre} DUT, a slight asymmetry in data can be observed for the DUT with \SI{100}{\micro\metre} thickness, likely originating from a residual misalignment of the DUT after the alignment procedure for data.
The distinct shape of the residuals is a result of the overlay of a broader peak from single-pixel clusters without correction and a narrow peak from two-pixel clusters with $\eta$-correction applied.

The simulated distributions have root mean square values of \SI{15.2}{\micro\metre} for the \SI{50}{\micro\metre} DUT and \SI{13.6}{\micro\metre} for the \SI{100}{\micro\metre} DUT, which are slightly larger than the values \SI{14.4}{\micro\metre} and \SI{12.8}{\micro\metre} obtained from data~\cite{AlipourTehrani}.
This can be attributed to the difference in cluster sizes described above.

The intrinsic resolution determined from simulation by quadratically subtracting the track resolution from the residual width is \SI{15.0}{\micro\metre} for the thinner and \SI{13.4}{\micro\metre} for the thicker DUT.

\paragraph{Monte Carlo truth information}

\begin{figure}[tbp]
  \centering
  \includegraphics[width=\columnwidth]{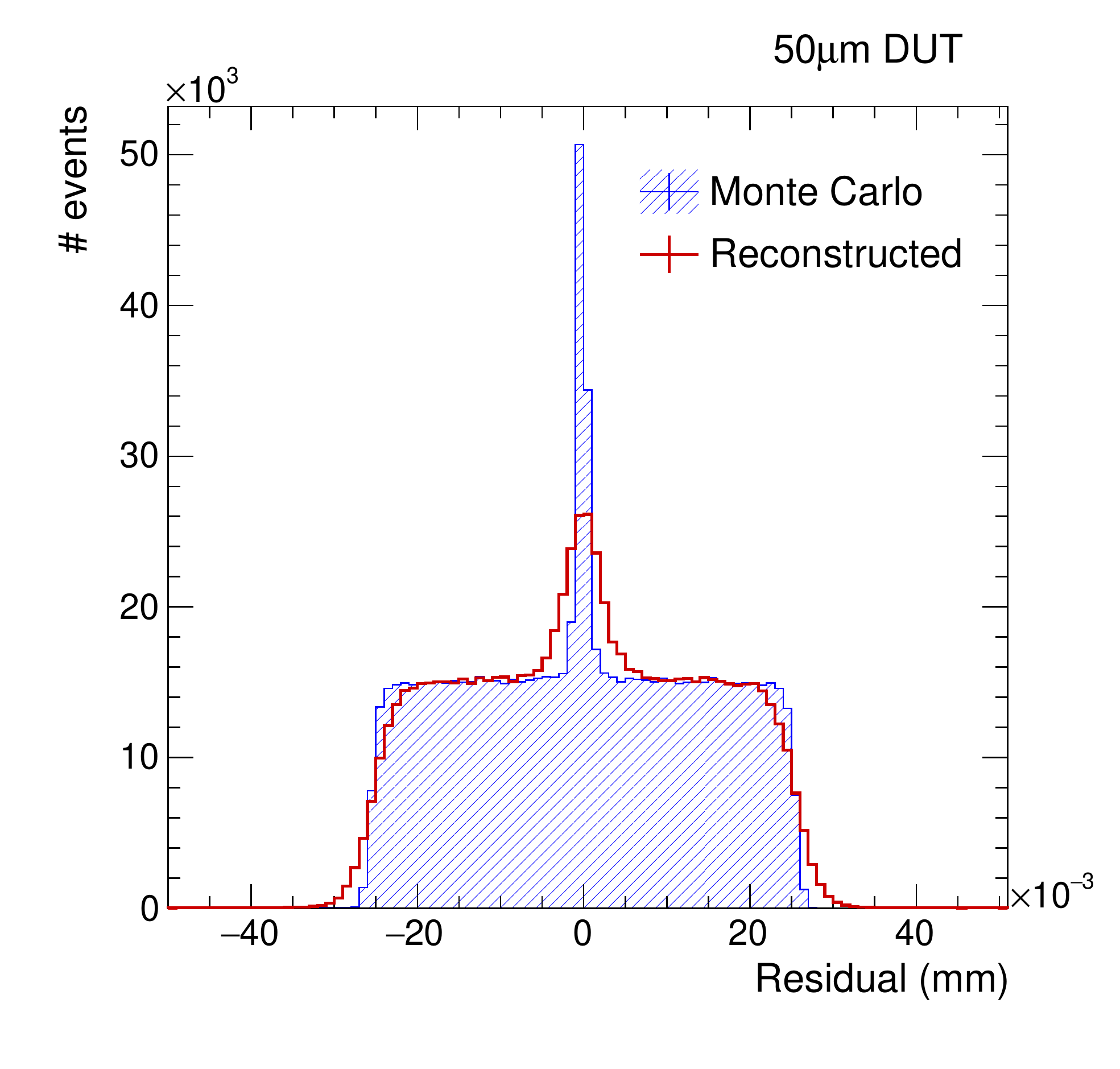}
  \caption{Comparison of track residuals calculated with the reconstructed track position and the recorded position of the Monte Carlo particle.}
  \label{fig:residuals:50MC}
\end{figure}

As described in Section~\ref{sec:framework_core}, the \apsq framework provides information on sensor entry and exit positions as well as type of the simulated Monte Carlo particle for each detector.
This allows for direct comparisons between the simulated particle and the reconstructed particle position after sensor and front-end simulation and track reconstruction.
An example for such a comparison is shown in Figure~\ref{fig:residuals:50MC}, which compares the residuals of the cluster position with the reconstructed telescope track and the Monte Carlo position for the \SI{50}{\micro\metre} thick DUT.
The additional smearing visible for the residual with telescope track position stems from the uncertainty of the track reconstruction in the beam telescope as shown in Figure~\ref{fig:telres}.


\section{Conclusions \& Outlook}
\label{sec:conclusion}
In this paper, the \apsq software for the simulation of silicon pixel detectors has been presented. \apsq is a lightweight and modular framework combining an elaborate simulation of interactions of particles and detector material with a detailed description of charge carrier motion and signal formation.

In addition to the framework's core, an initial set of simulation modules has been implemented and validated.
These comprise a module using Geant4 for the description of primary particle propagation and energy deposition, different methods for the propagation of charge carriers in silicon sensors and their coupling to the readout electronics, as well as a generic digitisation and various modules for data input and output.

The framework has been shown to perform very well when comparing data to the simulation result, adjusting only a few known parameters.
Similar track resolutions for the beam telescope have been measured for data and simulation, and both cluster size and charge of the DUT resemble the results obtained from data very closely.
The intrinsic resolution obtained from simulation is in agreement with data, and the shape of the residuals match very well.
Information about the primary particle position is available and can be used to gauge effects from charge transport, digitisation and track reconstruction on the detector performance.

Several extensions of the framework and additional modules are envisaged in order to further extend the functionality and to serve an even wider community.
This comprises a module for simulating the charge deposition of laser light, the simulation of transient effects via induced currents in the detector implants during charge carrier transport using Ramo weighting fields, and the effects of radiation damage in the silicon sensor.
Other ideas which might be realised within the framework are a digital front-end simulation for buffer-overflow studies and support for high-$Z$ sensor materials.


\section*{Acknowledgements}
\label{sec:acknowledgements}
This work was carried out in the framework of the CLICdp Collaboration.
This project has received funding from the European Union's Horizon 2020 Research and Innovation programme under Grant Agreement no. 654168.


\section*{References}
\bibliography{bibliography}

\begin{thebibliography}{10}
\expandafter\ifx\csname url\endcsname\relax
  \def\url#1{\texttt{#1}}\fi
\expandafter\ifx\csname urlprefix\endcsname\relax\def\urlprefix{URL }\fi
\expandafter\ifx\csname href\endcsname\relax
  \def\href#1#2{#2} \def\path#1{#1}\fi

\bibitem{allpix}
M.~Benoit, J.~Idarraga, \href{https://twiki.cern.ch:AllPix}{{The AllPix
  Simulation Framework -- Generic Geant4 implementation for pixel detectors}},
  accessed 6~2018.
\newline\urlprefix\url{https://twiki.cern.ch:AllPix}

\bibitem{kdetsim}
G.~Kramberger, \href{http://kdetsim.org/}{{KDetSim} - a simple way to simulate
  detectors}, accessed 6~2018.
\newline\urlprefix\url{http://kdetsim.org/}

\bibitem{pixelav}
M.~Swartz, \href{https://cds.cern.ch/record/687440}{{A Detailed Simulation of
  the CMS Pixel Sensor}}, Tech. Rep. CMS-NOTE-2002-027, CERN, Geneva (7 2002).
\newline\urlprefix\url{https://cds.cern.ch/record/687440}

\bibitem{clicdp-apsq-manual}
K.~Wolters, S.~Spannagel, D.~Hynds,
  \href{https://cds.cern.ch/record/2295206}{{User Manual for the Allpix$^2$
  Simulation Framework}}, Tech. Rep. CLICdp-Note-2017-006, CERN (12 2017).
\newline\urlprefix\url{https://cds.cern.ch/record/2295206}

\bibitem{apsq-website}
\href{https://cern.ch/allpix-squared/}{The {Allpix Squared} project}, accessed
  6~2018.
\newline\urlprefix\url{https://cern.ch/allpix-squared/}

\bibitem{tomlgit}
T.~Preston-Werner, \href{https://github.com/toml-lang/toml/}{{TOML} - {Tom's
  Obvious, Minimal Language}}, accessed 6~2018.
\newline\urlprefix\url{https://github.com/toml-lang/toml/}

\bibitem{apsq-repo}
\href{https://gitlab.cern.ch/allpix-squared/allpix-squared/}{The {Allpix
  Squared} software repository}, accessed 6~2018.
\newline\urlprefix\url{https://gitlab.cern.ch/allpix-squared/allpix-squared/}

\bibitem{iso-cpp14}
{International Organization for Standardization},
  \href{https://www.iso.org/standard/64029.html}{{Information technology --
  Programming languages -- C++}}, ISO/IEC 14882:2014, Geneva, Switzerland
  (2014).
\newline\urlprefix\url{https://www.iso.org/standard/64029.html}

\bibitem{doxygen}
D.~van Heesch, \href{https://www.stack.nl/~dimitri/doxygen/}{{Doxygen} -
  generate documentation from source code}, accessed 6~2018.
\newline\urlprefix\url{https://www.stack.nl/~dimitri/doxygen/}

\bibitem{geant4}
S.~Agostinelli, et~al., Geant4 -- a simulation toolkit, Nucl. Instr. Meth. A
  506~(3) (2003) 250 -- 303.
\newblock \href {http://dx.doi.org/10.1016/S0168-9002(03)01368-8}
  {\path{doi:10.1016/S0168-9002(03)01368-8}}.

\bibitem{geant4-2}
J.~Allison, et~al., Geant4 developments and applications, IEEE T. Nucl. Sci.
  53~(1) (2006) 270--278.
\newblock \href {http://dx.doi.org/10.1109/TNS.2006.869826}
  {\path{doi:10.1109/TNS.2006.869826}}.

\bibitem{geant4-3}
J.~Allison, et~al., Recent developments in {Geant4}, Nucl. Instr. Meth. A
  835~(Supplement C) (2016) 186 -- 225.
\newblock \href {http://dx.doi.org/10.1016/j.nima.2016.06.125}
  {\path{doi:10.1016/j.nima.2016.06.125}}.

\bibitem{pai}
J.~Apostolakis, et~al., {An implementation of ionisation energy loss in very
  thin absorbers for the GEANT4 simulation package}, Nucl. Instr. Meth. A 453
  (2000) 597--605.
\newblock \href {http://dx.doi.org/10.1016/S0168-9002(00)00457-5}
  {\path{doi:10.1016/S0168-9002(00)00457-5}}.

\bibitem{JACOBONI197777}
C.~Jacoboni, et~al., A review of some charge transport properties of silicon,
  Solid-State Electron. 20~(2) (1977) 77 -- 89.
\newblock \href {http://dx.doi.org/10.1016/0038-1101(77)90054-5}
  {\path{doi:10.1016/0038-1101(77)90054-5}}.

\bibitem{fehlberg}
E.~Fehlberg, \href{https://ntrs.nasa.gov/search.jsp?R=19690021375}{Low-order
  classical {Runge-Kutta} formulas with stepsize control and their application
  to some heat transfer problems}, NASA Technical Report NASA-TR-R-315, NASA
  (1969).
\newline\urlprefix\url{https://ntrs.nasa.gov/search.jsp?R=19690021375}

\bibitem{shockley}
W.~Shockley, Currents to conductors induced by a moving point charge, J. Appl.
  Phys. 9~(10) (1938) 635--636.
\newblock \href {http://dx.doi.org/10.1063/1.1710367}
  {\path{doi:10.1063/1.1710367}}.

\bibitem{ramo}
S.~Ramo, Currents induced by electron motion, Proc. IRE 27~(9) (1939) 584--585.
\newblock \href {http://dx.doi.org/10.1109/JRPROC.1939.228757}
  {\path{doi:10.1109/JRPROC.1939.228757}}.

\bibitem{root}
R.~Brun, F.~Rademakers, {ROOT} -- an object oriented data analysis framework,
  Nucl. Instr. Meth. A 389~(1–2) (1997) 81 -- 86, new Computing Techniques in
  Physics Research V.
\newblock \href {http://dx.doi.org/10.1016/S0168-9002(97)00048-X}
  {\path{doi:10.1016/S0168-9002(97)00048-X}}.

\bibitem{EUDET-2010-12}
I.~Rubinskiy,
  \href{http://www.eudet.org/e26/e28/e86887/e107460/EUDET-Memo-2010-12.pdf}{{EUTelescope.
  Offline track reconstruction and DUT analysis software}}, EUDET-Memo-2010-12.
\newline\urlprefix\url{http://www.eudet.org/e26/e28/e86887/e107460/EUDET-Memo-2010-12.pdf}

\bibitem{judith}
G.~McGoldrick, M.~\v{C}erv, A.~Gori\v{s}ek, Synchronized analysis of testbeam
  data with the judith software, Nucl. Instr. Meth. A 765 (2014) 140 -- 145.
\newblock \href {http://dx.doi.org/10.1016/j.nima.2014.05.033}
  {\path{doi:10.1016/j.nima.2014.05.033}}.

\bibitem{proteus}
\href{http://gitlab.cern.ch/unige-fei4tel/proteus}{The {Proteus} reconstruction
  framework}, accessed 6~2018.
\newline\urlprefix\url{http://gitlab.cern.ch/unige-fei4tel/proteus}

\bibitem{corry}
\href{http://gitlab.cern.ch/simonspa/corryvreckan}{The {Corryvreckan}
  reconstruction framework}, accessed 6~2018.
\newline\urlprefix\url{http://gitlab.cern.ch/simonspa/corryvreckan}

\bibitem{AlipourTehrani}
N.~Alipour~Tehrani, {Test-beam measurements and simulation studies of thin
  pixel sensors for the CLIC vertex detector}, Ph.D. thesis, CERN (2017).
\newblock \href {http://dx.doi.org/10.3929/ethz-b-000164813}
  {\path{doi:10.3929/ethz-b-000164813}}.

\bibitem{timepix3}
T.~Poikela, et~al., {Timepix3: a 65K channel hybrid pixel readout chip with
  simultaneous ToA/ToT and sparse readout}, J. Instr. 9~(05) (2014) C05013.
\newblock \href {http://dx.doi.org/10.1088/1748-0221/9/05/C05013}
  {\path{doi:10.1088/1748-0221/9/05/C05013}}.

\bibitem{Belau1983253}
E.~Belau, et~al., Charge collection in silicon strip detectors, Nucl. Instr.
  Meth. 214~(2--3) (1983) 253 -- 260.
\newblock \href {http://dx.doi.org/10.1016/0167-5087(83)90591-4}
  {\path{doi:10.1016/0167-5087(83)90591-4}}.

\end{thebibliography}

\end{document}